\documentclass[aps,prb,twocolumn,reprint,superscriptaddress]{revtex4-2}
\usepackage{graphicx}
\usepackage{amsmath}
\usepackage{amssymb}
\usepackage{colordvi}
\usepackage{mathrsfs}
\usepackage{bm}
\usepackage{verbatim}
\usepackage{dcolumn}
\usepackage{multirow}
\usepackage{rotating}
\usepackage{booktabs}
\usepackage{epsfig}
\usepackage{subfigure}
\usepackage{makecell}
\usepackage[colorlinks,allcolors=blue]{hyperref}

\begin{document}

\title{Electric-field-tunable topological phases in valley-polarized quantum anomalous Hall systems with inequivalent exchange fields}

\author{Shiyao Pan}
\thanks{These authors contributed equally in this work.}
\affiliation{Department of Physics, Fuzhou University, Fuzhou, Fujian 350108, China}

\author{Zeyu Li}
\thanks{These authors contributed equally in this work.}
\affiliation{International Center for Quantum Design of Functional Materials (ICQD), Hefei National Research Center for Physical Sciences at the Microscale, University of Science and Technology of China, Hefei 230026, China}

\author{Yulei Han}
\email[Correspondence author:~]{han@fzu.edu.cn}
\affiliation{Department of Physics, Fuzhou University, Fuzhou, Fujian 350108, China}


\begin{abstract}
Incorporating valley as a degree of freedom into quantum anomalous Hall systems offers a novel approach to manipulating valleytronics in electronic transport. Using the Kane-Mele monolayer as a concrete model, we comprehensively explore the various topological phases in the presence of inequivalent exchange fields and reveal the roles of the interfacial Rashba effect and external electric field in tuning topological valley-polarized states. We find that valley-polarized states can be realized by introducing Kane-Mele spin-orbit coupling and inequivalent exchange fields. Further introducing Rashba spin-orbit coupling and an electric field into the system can lead to diverse topological states, such as the valley-polarized quantum anomalous Hall effect with $\mathcal{C}=~\pm 1,\pm 2$ and valley-contrasting states with $\mathcal{C}=0$. Remarkably, different valley-polarized topological states can be continuously tuned by varying the strength and direction of the external electric field in a fixed system. Our work demonstrates the tunability of topological states in valley-polarized quantum anomalous Hall systems and provides an ideal platform for applications in electronic transport devices in topological valleytronics.
\end{abstract}
\maketitle

\section{Introduction}\label{sec1}
The quantum Hall effect, discovered in 1980, was the first topological phenomenon to exhibit quantized Hall conductivity and vanishing transverse resistance in the presence of a strong magnetic field~\cite{Klitzing2020}. The quantized Hall conductance can be denoted by $\mathcal{C} e^2/h$, where $\mathcal{C}$ is a topological number known as the Chern number and $e^2/h$ is the conductance quantum~\cite{TKNN}.  In 1988, Haldane proposed a model on a honeycomb lattice to realize nonzero quantized Hall conductance in the absence of the external magnetic field, known as the quantum anomalous Hall effect (QAHE)~\cite{Haldane,Ren2016,He2018,Chang_2023}. Unlike the traditional quantum Hall effect, the QAHE does not rely on strong magnetic fields but instead arises from the intrinsic magnetization of the materials, making it more feasible for applications in low-power electronics and topological quantum computing~\cite{Chang_2023}.
Over the past decade, the QAHE has been widely explored in various realistic systems~\cite{proposal1,proposal2,proposal3,proposal4,proposal5,proposal6,Yang2022_1, proposal7,proposal8,proposal9,Li2024,proposal10,proposal11,proposal12,proposal13,proposal14,proposal15,Li2023,proposal16,proposal17,proposal18,proposal19,Wang_2023,proposal20,QAH-Ctunable1,QAH-graphene_CrI3,QAH-Ctunable2,QAH-Ctunable3}.

In a honeycomb lattice, two inequivalent valleys, $K$ and $K'$, exist at the corner of the first Brillouin zone due to the presence of $C_3$ rotation symmetry. This provides a new degree of freedom for controlling electronic states, known as valleytronics. By breaking inversion symmetry, the system opens a band gap in the two valleys with opposite Berry curvatures, resulting in the valley Hall effect~\cite{QVH1,QVH2,Zheng_2023,QVH3,Guo_2024,QVH_C,Zeng_2022,Tang_2024,Xun_2024,Li_2023_2,Li_2024}, where electrons propagate in opposite directions for the two valleys.

Incorporating valley degrees of freedom into QAHE systems not only enhances the robustness of valley transport against disorders but also increases the electronic tunability of QAHE. This new topological phase, known as valley-polarized QAHE, was first proposed in monolayer silicene with a uniform exchange field~\cite{proposal20} and has since been explored in various systems~\cite{VP-QAH-disorder,VP-QAH-electronic,VP-ferri, half-H-Bi bilayer,VP-H-Bilayer,VP-MX2,VP-MoSiX,Valley-symmetry-broken,VP+QAH1,VP+QAH2,VP-QAH-Ctunable3,Guo_2023,Li_2023}. However, these systems often require intricate artificial decoration or possess small band gaps, necessitating the search for new systems with larger band gaps. The discovery of large gap Kane-Mele type topological insulators, such as the Pt$_2$HgSe$_3$ monolayer, presents an opportunity to realize valley-polarized QAHE with large band gap~\cite{PHS experiment,PHS1,PHS2}. By introducing inequivalent magnetic moments in Hg atoms and leveraging interfacial Rashba effect, valley-polarized QAHE can be achieved in the monolayer Pt$_2$HgSe$_3$ through van der Waals coupling to different magnetic insulators~\cite{Liu2021, Zhu2023}. The topological phase in these systems is influenced by competing factors such as inequivalent  magnetic exchange fields, spin-orbit coupling (SOC) and electric field, with the underlying mechanisms yet to be fully understood. Moreover, tuning the topological states in QAHE systems is challenging due to the fixed Chern number in a specific sample, significantly limiting the practical application of QAHE-based electronic devices.

In this work, we theoretically investigate the topological phases of valley-polarized QAHE systems in the presence of inequivalent magnetic exchange fields. We thoroughly consider the effects of inequivalent exchange fields, Rashba spin-orbit coupling, and external electric fields. We first reveal the distinct roles of uniform and inequivalent exchange fields, demonstrating that valley-polarized states can be achieved by considering both Kane-Mele spin-orbit coupling and inequivalent exchange fields. Subsequently, we present the topological phase diagram in the parameter space of interfacial Rashba spin-orbit coupling and external electric fields. By selecting five representative points, we explore topological properties such as Berry curvatures, anomalous Hall conductivity, density of states, and orbital magnetic properties. Notably, we find that the electric field is a crucial tool for tuning the topological phase in a fixed valley-polarized system, and the large orbital magnetic moments can induce orbital-magnetization-based optical Kerr effects. Our work provides an effective method for controlling topological states in valley-polarized QAHE systems.

\section{Model and Method}
\begin{figure}
  \centering
  \includegraphics[width=8.5cm]{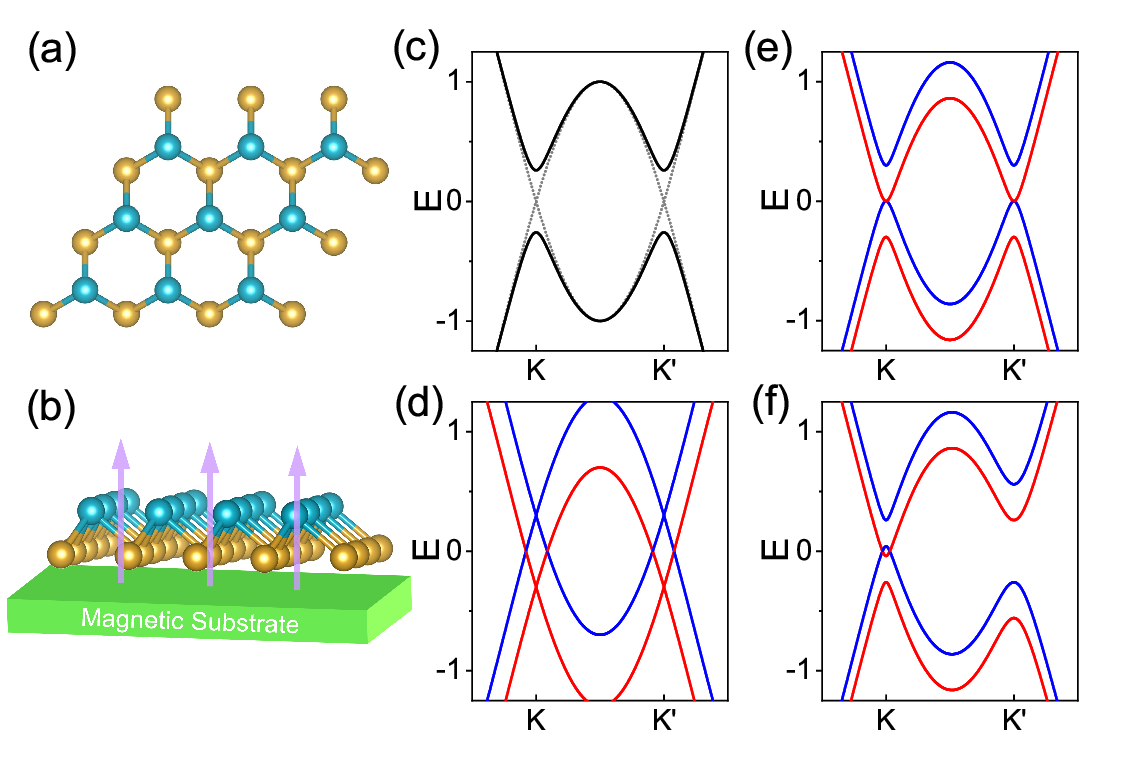}
  \caption{(a)-(b) Schematic diagram of the system with a buckled honeycomb lattice, depicted by yellow and cyan spheres. The purple arrows indicate the magnetization direction of the substrate. (c)-(f) Band structures of the system with different parameters: (c) $t_{\mathrm{SO}}=0.05$, $m_{1,2}=0$, (d) $t_{\mathrm{SO}}=0$, $m_{1,2}=0.3$, (e) $m_{1}=0.3$, $t_{\mathrm{SO}}=m_2=0$, (f) $t_{\mathrm{SO}}=0.05$, $m_{1}=0.3$, $m_2=0$. Black, red, and blue colors represent spin-degenerate, spin-up and spin-down states, respectively. The dashed line in (c) denotes the Dirac cones in the absence of spin-orbit coupling.}\label{Fig1}
\end{figure}

To explore the tunability of topological phases in valley-polarized QAHE systems with inequivalent magnetic exchange fields, we consider a buckled honeycomb lattice coupled to a magnetic substrate, as displayed in Figs.~\ref{Fig1}(a)-\ref{Fig1}(b). Due to the different distance between the buckled sublattices and the magnetic substrate, the magnetic proximity effect can induce inequivalent exchange fields in the honeycomb lattice~\cite{Liu2021}. Based on the Kane-Mele model, the tight-binding Hamiltonian can be written as~\cite{proposal15}:
\begin{eqnarray}\label{eq:Ham}
H & = & t\sum_{\langle ij \rangle}c_{i}^{\dagger}c_{j}+i t_{\mathrm{SO}}\sum_{\langle\langle ij \rangle\rangle}v_{ij}s_{z}c_{i}^{\dagger}c_{j}+i t_{\mathrm{R}}\sum_{\langle ij \rangle}(\mathbf{s}\times \mathbf{\hat{d}}_{ij})^{z}c_{i}^{\dagger}c_{j} \notag\\
&& +\sum_{i}m_{\alpha}s_{z} c_{i}^{\dagger}c_{i}+\sum_{i} \delta_{\alpha}c_{i}^{\dagger}c_{i},
\end{eqnarray}
where $c^{\dagger}_{i}$($c_{i}$) is the creation (annihilation) operator for electrons on site $i$, $\langle ...\rangle$ ($\langle\langle ...\rangle\rangle$) sums over for the (next-)nearest neighbor sites, and $\mathbf{s}=(s_x,s_y,s_z)$ represents the Pauli matrice for spin degrees of freedom. The first term is the nearest-neighbor hopping in the honeycomb lattice with hopping energy $t$. The second term represents the Kane-Mele SOC with an amplitude of $t_{\mathrm{SO}}$, and $v_{ij}=\pm 1$ depends on the clockwise/counterclockwise orientation of the next-nearest-neighbor hopping path. The third term describes the Rashba SOC, with strength $t_\mathrm{R}$, arising from the structural inversion asymmetry due to the presence of magnetic substrate. The fourth term represents the inequivalent exchange fields induced in A/B sublattices with strengths $m_1$ and $m_2$, respectively, which breaks the time-reversal symmetry of the system. The last term denotes the sublattice staggered potential induced by an out-of-plane electric field in this buckled system, with a strength of $\delta$, where $\delta=\delta_{\mathrm{A}}=-\delta_{\mathrm{B}}$. Here, we set $t=1$ as the unit of energy and $t_\mathrm{SO}=0.05$.
We neglect the influence of the external electric field on the Rashba SOC since that $t_\mathrm{R}$ induced by a perpendicular external electric field is weak in two-dimensional Dirac systems~\cite{Kane_2005}.

We also investigate the topological properties of the system, i.e., the Chern numbers, which can be calculated by~\cite{TKNN}:
\begin{eqnarray}
  \mathcal{C} &=& \frac{1}{2\pi}\sum_{n}\int_{BZ} d^2 \mathbf{k}\Omega_n(\mathbf{k}),\\
  \Omega_n(\mathbf{k})&=&-2\sum_{n'\neq n}\frac{\mathrm{Im}\langle u_{n\mathbf{k}}|v_x|u_{n'\mathbf{k}}\rangle \langle u_{n'\mathbf{k}}|v_y|u_{n\mathbf{k}}\rangle}{(E_{n'}-E_{n})^2}  ,
\end{eqnarray}
where $\Omega_n(\mathbf{k})$ represents the Berry curvature and the summation is over all occupied bands in the first Brillouin zone, $u_n(\mathbf{k})$ is the periodic part of the Bloch wave functions, $v_{x,y}$ is the velocity operator. We also study the orbital-related properties, i.e., the orbital magnetic moment $m_n(\mathbf{k})$ and orbital magnetization $M_z(\mu)$, which can be respectively calculated by~\cite{Xiao2010}:
\begin{eqnarray}
 m_n(\mathbf{k})&=&\frac{e}{\hbar} \sum_{n'\neq n}\frac{\mathrm{Im}\langle u_{n\mathbf{k}}|v_x|u_{n'\mathbf{k}}\rangle \langle u_{n'\mathbf{k}}|v_y|u_{n\mathbf{k}}\rangle}{E_{n'}-E_{n}},\\
M_z(\mu)&=&\sum_n \int d^2 \mathbf{k} \left[ m_n(\mathbf{k}) + \frac{e}{\hbar} \Omega_n(\mathbf{k}) (\mu - E_{n \mathbf{k}}) \right],
\end{eqnarray}
where $E_n (\mathbf{k})$ is the $n$-th eigenvalue at $\mathbf{k}$ and $\mu$ represents the chemical potential.

\section{Results and Discussions}
\subsection{The influence of inequivalent exchange fields}
We first demonstrate the differences between uniform and inequivalent exchange fields on the band structures. In the absence of an exchange field, as displayed in Fig.~\ref{Fig1}(c), the presence of Kane-Mele SOC induces a band gap of $6\sqrt{3}t_{\mathrm{SO}}$ around $K/K'$ valleys. When a uniform exchange field is considered, as depicted in Fig.~\ref{Fig1}(d), a spin splitting of size $2m$ exists in the band, while the linear dispersions around $K/K'$ valleys are still preserved. In contrast, when only the inequivalent exchange field is considered, e.g., $m_1=0.3$ and $m_2=0.0$, as shown in Fig.~\ref{Fig1}(e), a spin splitting of size $m_1$ appears around the two valleys, and the linear Dirac cones transform into parabolic dispersion. This indicates the different roles of uniform and inequivalent exchange fields in determining the band properties. It is important to note that the aforementioned three cases preserve valley degeneracy, i.e., the band energies of the two valleys are degenerate due to preserved $\mathcal{PT}$ symmetry.
To break the valley degeneracy, both Kane-Mele SOC and inequivalent exchange fields must be simultaneously included in the system, as demonstrated in Fig.~\ref{Fig1}(f), where a large valley splitting appears. Specifically, the four band energies at the $K$ valley are $\pm 3\sqrt{3}t_{\mathrm{SO}}$ and $\pm 3\sqrt{3}t_{\mathrm{SO}}\mp m_1$, whereas the band energies at the $K'$ valley are $\pm 3\sqrt{3}t_{\mathrm{SO}}$ and $\pm 3\sqrt{3}t_{\mathrm{SO}}\pm m_1$, leading to a spin and valley splitting of $m_1$.

\begin{figure}
  \centering
  \includegraphics[width=8.5cm]{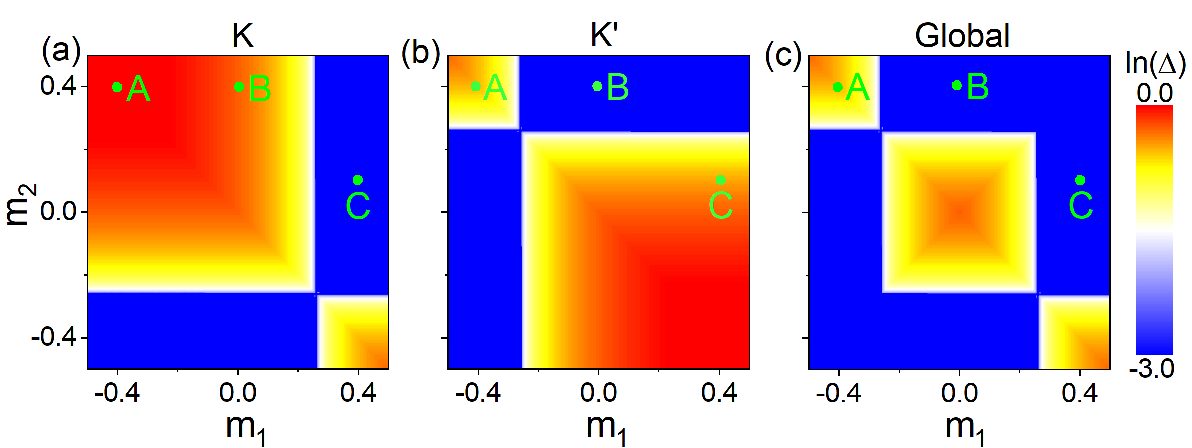}
  \caption{Phase diagrams of band gaps $\Delta$ in ($m_1$, $m_2$) space for (a) the $K$ valley, (b) the $K'$ valley and (c) bulk systems. The blue color indicates a metallic phase. Three different points $A$, $B$ and $C$ are labeled in green, with ($m_1$, $m_2$) = (0.4, -0.4), (0.0, 0.4), and (0.4, 0.1), respectively. $t_\mathrm{SO}$ is included.}\label{Fig2}
\end{figure}

From the above analysis, we can find that the band gap is determined by the strength of $t_{\mathrm{SO}}$ and $m_{1,2}$. Figure~\ref{Fig2} displays the phase diagram of band gaps for valley $K/K'$ and bulk in the exchange fields ($m_1$, $m_2$) space. The phase boundaries separating the insulating and metallic states are clearly observable, specifically at $m_{1,2}=\pm 3\sqrt{3} t_{\mathrm{SO}}$.
For the local gap around valley $K$, as demonstrated in Fig.~\ref{Fig2}(a), the phase boundaries are $m_1=-m_2=3\sqrt{3} t_{\mathrm{SO}}$. Conversely, the phase boundaries for the local gap around valley $K'$ are $m_1=-m_2=-3\sqrt{3} t_{\mathrm{SO}}$, due to the valley-dependent Kane-Mele SOC [see Fig.~\ref{Fig2}(b)]. The bulk band gap is evaluated by the band alignment of the two valleys, as shown in Fig.~\ref{Fig2}(c), where three insulating regions can be identified. Since reversing the sign of $m_{1,2}$ only exchanges the spin-up and spin-down bands without affecting the shape of band structures, the phase diagrams are symmetric along the line of $m_1=-m_2$.
Moreover, compared to the uniform exchange field with $m_1=m_2$, the presence of the inequivalent exchange field not only induces valley splitting but also provides a rich phase diagram that can be further tuned by external fields.

\begin{figure}
  \centering
  \includegraphics[width=8.5cm]{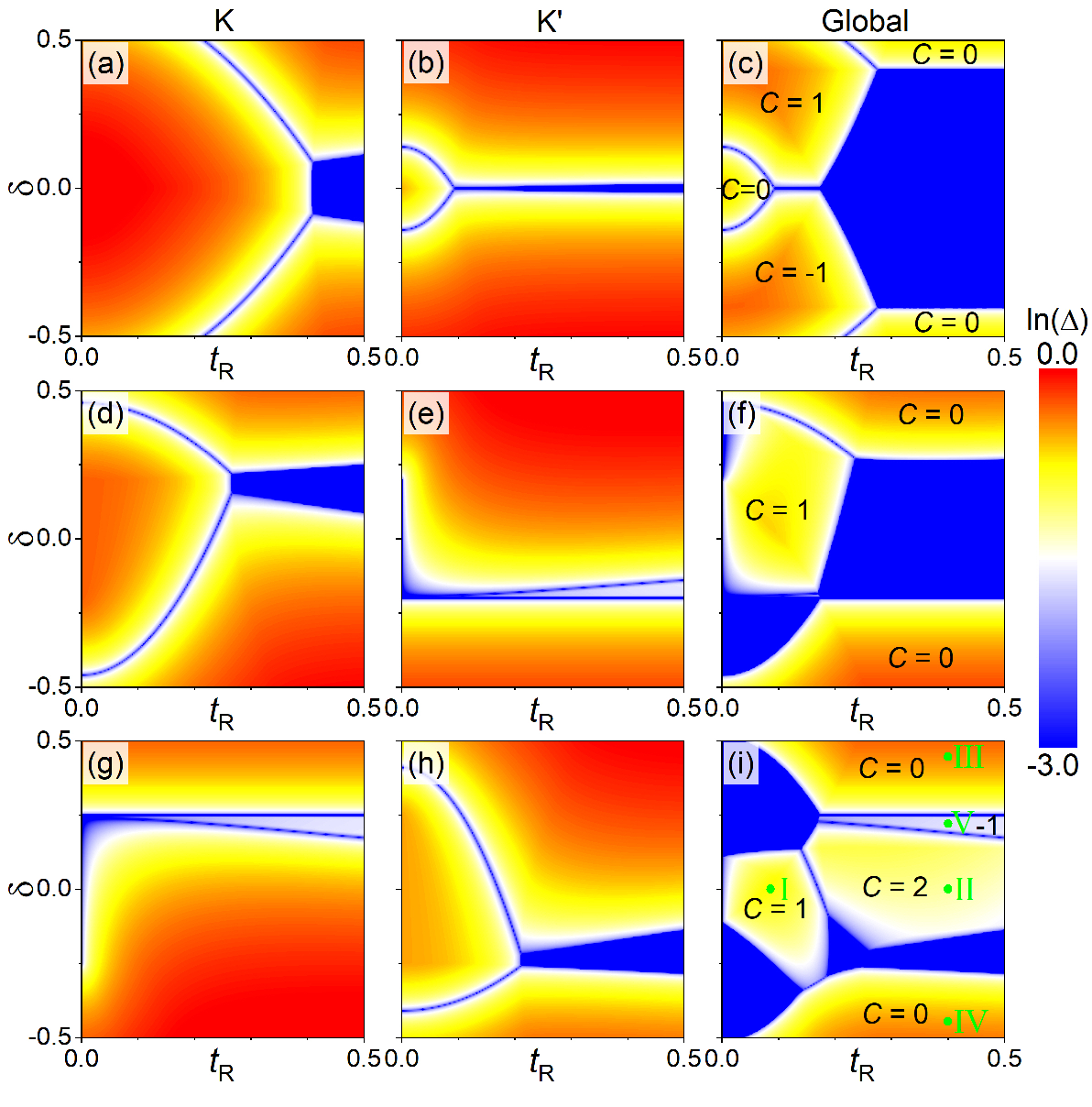}
  \caption{Phase diagrams of band gaps $\Delta$ in ($t_\mathrm{R}$, $\delta$) space for (a)-(c) point $A$, (d)-(f) point $B$, and (g)-(i) point $C$. The blue color indicates a metallic phase. The first, second, and third columns correspond to band gaps around the $K$ valley, $K'$ valley and bulk system, respectively. The Chern number for each insulating region is labeled in the third column. Five representative points \uppercase\expandafter{\romannumeral1}-\uppercase\expandafter{\romannumeral5} are labeled in (i), with ($t_\mathrm{R}$, $\delta$) = (0.1, 0.0), (0.4, 0.0), (0.4, 0.45), (0.4, -0.45), and (0.4, 0.21), respectively. }\label{Fig3}
\end{figure}

\begin{figure*}
  \centering
  \includegraphics[width=17.0cm]{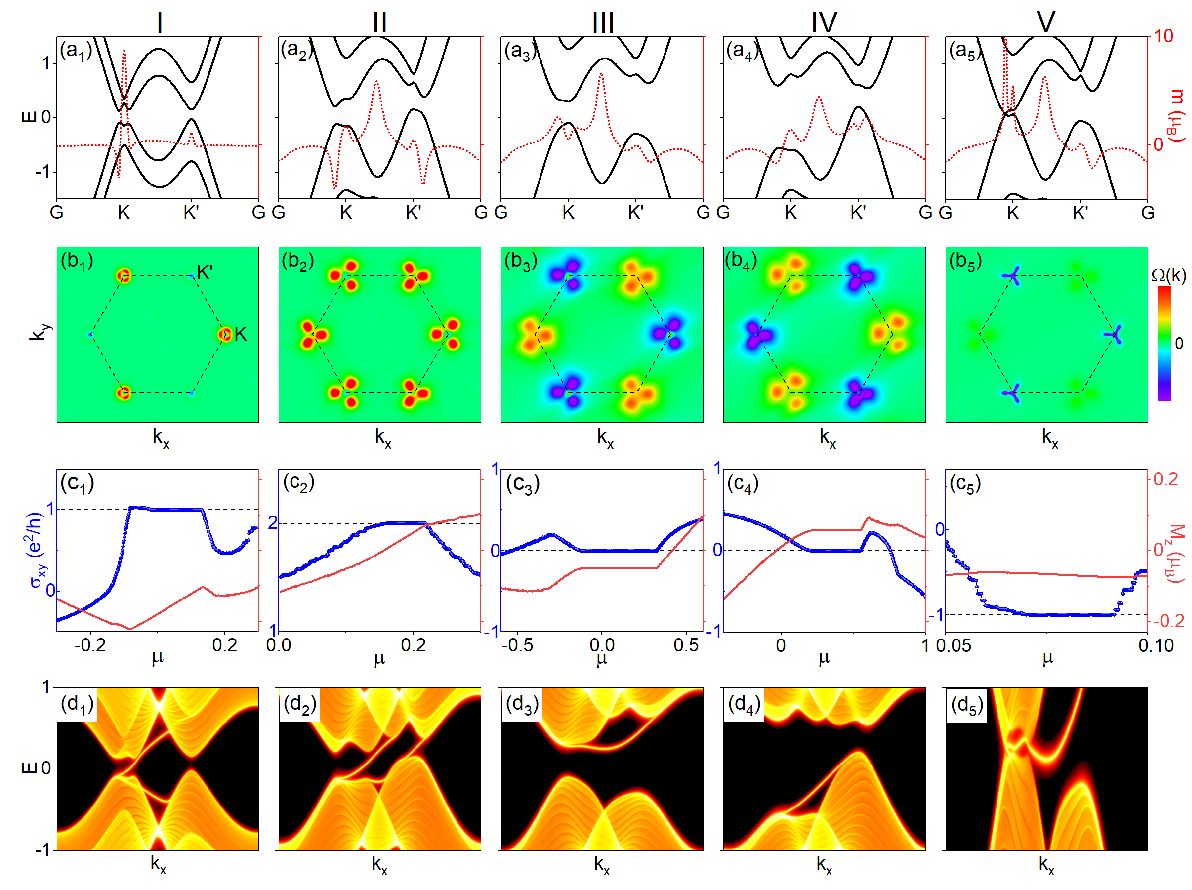}
  \caption{The topological properties of the five representative points \uppercase\expandafter{\romannumeral1}-\uppercase\expandafter{\romannumeral5} labeled in Fig.~\ref{Fig3}(i). (a$_1$)-(a$_5$): Band structures and orbital magnetic moments $m(k)$ for the occupied bands along high-symmetry lines. (b$_1$)-(b$_5$): Distributions of Berry curvatures $\Omega (k)$ for the occupied bands in the first Brillouin zone, indicated by black dashed lines. Red (blue) color denotes the positive (negative) $\Omega (k)$.  (c$_1$)-(c$_5$): Anomalous Hall conductivity $\sigma_{xy}$ and orbital magnetization $M_z$ per unit cell as a function of Fermi level. A typical lattice constant of 4 \AA~is used to evaluate $M_z$. (d$_1$)-(d$_5$): Density of states of semi-infinite zigzag nanoribbons. }\label{Fig4}
\end{figure*}

\subsection{The influence of Rashba SOC and external electric field}
According to the local gap $\Delta_{K/K'}$ around valley $K/K'$, we choose three different points, i.e., $A$ ($\Delta_K\neq 0$, $\Delta_{K'}\neq 0$), $B$ ($\Delta_K\neq 0$, $\Delta_{K'}= 0$) and $C$ ($\Delta_K= 0$, $\Delta_{K'}\neq 0$), to explore the influence of interfacial Rashba effect and vertical electric field on the topological properties of the system.
The phase diagrams of band gaps $\Delta$ in the ($t_\mathrm{R}$, $\delta$) space for these three points are summarized in Fig.~\ref{Fig3}.

As displayed in Figs.~\ref{Fig3}(a)-\ref{Fig3}(c) for point $A$, increasing of the Rashba SOC strength $t_{\mathrm{R}}$ drives the two valleys and bulk system into metallic states in the absence of a vertical  electric field ($\delta = 0$). Conversely, in the absence of Rashba SOC, applying a positive electric field leads to a local gap closing and reopening around valley $K'$, driving the system from a trivial insulator with $\mathcal{C}=0$ to a valley-polarized QAHE state with $\mathcal{C}=1$. If the direction of the vertical electric field is reversed, the sign of Chern number is also reversed, resulting in a topological phase transition from $\mathcal{C}=0$ to $\mathcal{C}= -1$.
Due to the bulk-boundary correspondence, the sign of Chern number determines the direction of electrons movement along the sample edges. Thus, tuning the direction of the electric field is a feasible method to control the electronic transport in valley-polarized QAHE systems.
For a large electric field strength, e.g., $\delta=0.5$, increasing $t_{\mathrm{R}}$ induces a band gap closing and reopening around the $K$ valley and drives the system from a $\mathcal{C}=1$ QAHE state to another $\mathcal{C}=0$ state. We should note that the above two $\mathcal{C}=0$ insulating states, i.e., the left middle part and the right upper part in Fig.~\ref{Fig3}(c), are topologically inequivalent. For the left middle part, e.g., ($t_\mathrm{R}$, $\delta$) = (0, 0), the Berry curvatures vanish throughout the Brillouin zone. In contrast, for the right upper part, e.g., ($t_\mathrm{R}$, $\delta$) = (0.4, 0.5), the nonzero Berry curvatures around the two valleys have opposite signs, leading to a zero Chern number. These valley-polarized $\mathcal{C}=0$ states can be utilized for valley-contrasting electronic transport, where the current can be valley-selective.

We then turn to point $B$, as demonstrated in Figs.~\ref{Fig3}(d)-\ref{Fig3}(f), the phase diagrams for valley $K$ in Fig.~\ref{Fig3}(d) resemble those in Fig.~\ref{Fig3}(a) because both points $A$ and $B$ are in the same insulating region shown in Fig.~\ref{Fig2}(a).
In the absence of $t_\mathrm{R}$ and $\delta$, the band around valley $K'$ is metallic due to the band crossing of conduction and valence bands. When a small Rashba SOC is introduced, a local gap opens at the band crossing points, transforming the system from metallic to a valley-polarized QAHE insulator with $\mathcal{C}=1$, as depicted in Figs.~\ref{Fig3}(e)-\ref{Fig3}(f).
When a electric field is applied, we can find a critical phase boundary at $\delta = -m_2$ in Fig.~\ref{Fig3}(e) for valley $K'$. Further increasing the negative electric field strength drives the system into an insulating state without band crossing.
Due to the band alignment of the two valleys, three insulting regions can be identified in Fig.~\ref{Fig3}(f). Notably, the two $\mathcal{C}=0$ regions exhibit opposite nonzero Berry curvature distributions around the two valleys, which can be utilized in the valley Hall effect.

For point $C$, as shown in Figs.~\ref{Fig3}(g)-\ref{Fig3}(h), the phase diagram of the $K$ ($K'$) valley is similar to that of the $K'$ ($K$) valley for point $B$ in Figs.~\ref{Fig3}(d)-\ref{Fig3}(e), due to the opposite local gap states of the two valleys at points $B$ ($\Delta_K\neq 0$, $\Delta_{K'}= 0$) and $C$ ($\Delta_K= 0$, $\Delta_{K'}\neq 0$) [see Fig.~\ref{Fig2}]. Interestingly, this exchange of phase diagrams between the two valleys results in a distinct phase diagram for the overall system as displayed in Fig.~\ref{Fig3}(i), where five insulating regions with different Chern numbers can be observed.
The difference in phase diagrams between points $B$ and $C$ originates from the inclusion of $m_2$, which alters the local gap around $K$/$K’$ valley and leads to different band alignments in the presence of the electric field and Rashba SOC~\cite{SM}.
These findings suggest that the inequivalent exchange fields, interfacial Rashba effect as well as the electric field play crucial roles in determining the topological phases of valley-polarized QAHE systems.
In particular, the external electric field emerges as a important tool for tuning experimentally the topological states within a fixed system, and we will elucidate it below.

\subsection{Electric-field tunable topological properties}
To demonstrate the electric-field-tunable topological properties, we analyze five representative points \uppercase\expandafter{\romannumeral1}-\uppercase\expandafter{\romannumeral5} labeled in Fig.~\ref{Fig3}(i), where points \uppercase\expandafter{\romannumeral2}-\uppercase\expandafter{\romannumeral5} are characterized by different electric field strengths but share the same $t_\mathrm{R}$.

For point  \uppercase\expandafter{\romannumeral1} with $\mathcal{C}=1$, as shown in Figs.~\ref{Fig4}(a$_1$)-\ref{Fig4}(b$_1$), one can observe a band inversion around the $K$ valley, leading to a large peak in orbital magnetic moments of approximately 9 $\mu_{\mathrm{B}}$ and large positive Berry curvature distributions for valence bands. In contrast, the absence of band inversion around the $K'$ valley results in negligible orbital magnetic moment $m(k)$ and Berry curvatures. Therefore, the Berry curvatures are primarily contributed by the valley $K$, causing the anomalous Hall conductivity $\sigma_{xy}$ to approach the quantized value of $e^2/h$ within the bulk gap, as displayed in Fig.~\ref{Fig4}(c$_1$), accompanied by a large orbital magnetization $M_z$ of approximately -0.22$\mu_\mathrm{B}$ per unit cell. This $\mathcal{C}=1$ topological state is further confirmed by the density of states of semi-infinite zigzag nanoribbons, as shown in Fig.~\ref{Fig4}(d$_1$), where a gapless state exists within the band gap.

For point \uppercase\expandafter{\romannumeral2} with $\mathcal{C}=2$, as displayed in Fig.~\ref{Fig4}(a$_2$), band inversions are observed around both valleys, each with orbital magnetic moments of approximately -4 $\mu_{\mathrm{B}}$. Additionally, a large $m(k)$ peak of about 6 $\mu_{\mathrm{B}}$ is present at the high-symmetry point $M$. The positive Berry curvature distributions appear around both valleys, resulting in a quantized anomalous Hall conductivity of $2~e^2/h$ within the bulk gap, as shown in Figs.~\ref{Fig4}(b$_2$)-\ref{Fig4}(c$_2$). From Fig.~\ref{Fig4}(d$_2$), one can also observe that two conducting channels exists within the band gap.

For point \uppercase\expandafter{\romannumeral3} with $\mathcal{C}=0$, a large $m(k)$ peak of about 7 $\mu_{\mathrm{B}}$ exists at the high-symmetry point $M$ in Fig.~\ref{Fig4}(a$_3$). The valley-contrasting Berry curvature distributions, illustrated in Fig.~\ref{Fig4}(b$_3$), reveal negative contributions from the $K$ valley and positives contributions from the $K'$ valley, resulting in a net zero $\sigma_{xy}$ and the absence of gapless states within the bulk gap, as demonstrated in Figs.~\ref{Fig4}(c$_3$) and \ref{Fig4}(d$_3$). Since the slope $dM_{z}/d\mu$ within the gap is proportional to $\mathcal{C}$~\cite{Ceresoli_2006,SM}, the orbital magnetization remains constant within the bulk gap at -0.05 $\mu_\mathrm{B}$ per unit cell.
Moreover, both the valence band maximum and conduction band minimum originate from the $K$ valley, leading to a large valley splitting of valence band about $0.2t$ and a large $\sigma_{xy}$ outside the bulk gap.
If the direction of external electric field is reversed, as shown for point \uppercase\expandafter{\romannumeral4} in Figs.~\ref{Fig4}(a$_4$)-\ref{Fig4}(d$_4$), the bands near $E=0$ are primarily contributed by the $K'$ valley, resulting in a significant valley splitting of the valence band about $0.76t$. Conversely, the $K$ and $K'$ valleys have positive and negative contributions to the Berry curvatures, respectively, inducing a net zero $\sigma_{xy}$ and an orbital magnetization of 0.06 $\mu_\mathrm{B}$ per unit cell within the band gap. Consequently, the nonzero $\sigma_{xy}$ near the band gap originates from carriers around the $K'$ valley.

Interestingly, as shown in Fig.~\ref{Fig3}(i), it is important to note that the region corresponding to point \uppercase\expandafter{\romannumeral5} with $\mathcal{C}=-1$ separates the regions between points \uppercase\expandafter{\romannumeral2} and \uppercase\expandafter{\romannumeral3} with $\mathcal{C}=2$ and $\mathcal{C}=0$, respectively. Figures~\ref{Fig4}(a$_5$)-\ref{Fig4}(d$_5$) illustrate the topological properties of point \uppercase\expandafter{\romannumeral5}. The bands near the Fermi level are primarily contributed by the $K'$ valley, with a valley splitting of the valence band about $0.12t$. The orbital magnetic moment reaches a maximum of about 28 $\mu_\mathrm{B}$ near the $K$ valley. The Berry curvatures exhibit large negative peaks around the $K$ valley, resulting in a quantized $\sigma_{xy}=-e^2/h$ within the bulk gap and a gapless state connecting the valence and conduction bands around the $K$ valley. The orbital magnetization is negative near the Fermi level, with a value of about -0.09 $\mu_\mathrm{B}$ per unit cell at the Fermi level.

From the above analysis, it is evident that different insulating regions exhibit distinct valley-polarized topological properties, including valley-polarized QAHE with $\mathcal{C}=\pm 1,~2$ and valley-polarized states with valley-contrasting Berry curvatures. The large orbital magnetic moment can also induce an orbital-magnetization-based optical Kerr effect in these valley-polarized systems. Remarkably, the regions for points \uppercase\expandafter{\romannumeral2}-\uppercase\expandafter{\romannumeral5} can be achieved by continuously tuning the strength and direction of the external electric field, indicating that different valley-polarized topological states can be realized in a single fixed system.

\section{Summary}
In this work, we systematically investigate various topological phases in a Kane-Mele monolayer in the presence of inequivalent exchange fields, and identify the electric-field-tunable valley-polarized states in this system.
First, we demonstrate that valley-polarized states can be induced by including both the Kane-Mele SOC and inequivalent exchange fields, and we obtain the phase diagram in sublattice-inequivalent exchange field space.
Next, we consider the roles of the interfacial Rashba effect and external electric field, exploring the topological phase diagrams in these parameter spaces under three different conditions.
By choosing five representative points, we thoroughly analyze the topological properties, including Berry curvatures, anomalous Hall conductivity, density of states of semi-infinite zigzag nanoribbons, and orbital magnetic properties.
We find different insulating regions with various valley-polarized topological states, including valley-polarized QAHE with $\mathcal{C}=\pm 1,\pm 2$ and valley-contrasting states with $\mathcal{C}=0$. Remarkably, these valley-polarized topological states can be continuously tuned by adjusting the strength and direction of the external electric field within a fixed system.
In addition, since the size of the initial band gap is proportional to the strength of intrinsic SOC $t_\mathrm{SO}$, reducing $t_\mathrm{SO}$ is a feasible way to diminish the required strengths of the exchange fields, electric field, and Rashba SOC. 

Moreover, these electrically tunable valley-polarized topological states can be realized by introducing magnetism to Kane-Mele topological insulators such as silicene, germanene, stanene~\cite{proposal15,Liu_2011}, and Pt$_2$HgSe$_3$ family of materials~\cite{PHS experiment,PHS1,PHS2}, through van der Waals stacking with two-dimensional magnetic insulators~\cite{Liu2021, Zhu2023}. As classical counterparts of QAHE and Kane-Mele model, photonic and sonic crystals are also promising platforms to realize these topological states~\cite{Wang_2015,Mousavi_2015,Khanikaev_2017,Mittal_2019}.
Our work reveals the crucial role of the electric field in tuning valley-polarized topological systems and will inspire further exploration on the application of electronic transport devices in topological valleytronics.

\begin{acknowledgments}
This work was financially supported by the National Natural Science Foundation of China (No. 12004369), Natural Science Foundation of Fujian Province (No. 2022J05019), China Postdoctoral Science Foundation (No. 2023M733411 and 2023TQ0347), and the Education and Research fund for Young Teachers of Fujian Province (No. JAT210014).
\end{acknowledgments}


\begin{thebibliography}{99}
\bibitem{Klitzing2020}
K. von Klitzing, T. Chakraborty, P. Kim, V. Madhavan, X. Dai, J. McIver, Y. Tokura, L. Savary, D. Smirnova, A. Maria Rey, C. Felser, J. Gooth, and X. Qi, \text{Nat. Rev. Phys.} \textbf{2}, 397 (2020).

\bibitem[Thouless(1982)]{TKNN}
D. J. Thouless, M. Kohmoto, M. P. Nightingale, and M. denNijs, \text{Phys. Rev. Lett.} \textbf{49}, 405 (1982).

\bibitem[Haldane(1988)]{Haldane}
F. D. M. Haldane, \text{Phys. Rev. Lett.} \textbf{61}, 2015 (1988).

\bibitem{Chang_2023}
C.-Z. Chang, C.-X. Liu, and A. H. MacDonald, \text{Rev. Mod. Phys.} \textbf{95}, 011002 (2023).

\bibitem[He(2018)]{He2018}
K. He, Y. Wang, and Q.-K. Xue, \text{Annu. Rev. Condens. Matter Phys.} \textbf{9}, 329 (2018).

\bibitem[Ren(2016)]{Ren2016}
Y. Ren, Z. Qiao, and Q. Niu, \text{Rep. Prog. Phys.} \textbf{79}, 066501 (2016).

\bibitem[Onoda(2003)]{proposal1}
M. Onoda, and N. Nagaosa, \text{Phys. Rev. Lett.} \textbf{90}, 206601 (2003).

\bibitem[Liu(2008)]{proposal2}
C. Liu, X. Qi, X. Dai, Z. Fang, and S.-C. Zhang,  \text{Phys. Rev. Lett.} \textbf{101}, 146802 (2008).

\bibitem[Wu(2008)]{proposal3}
C. Wu, \text{Phys. Rev. Lett.} \textbf{101}, 186807 (2008).

\bibitem[Yu(2010)]{proposal4}
R. Yu, W. Zhang, H. Zhang, S.-C. Zhang, X. Dai, and Z. Fang, \text{Science} \textbf{329}, 61 (2010).

\bibitem[Qiao(2010)]{proposal5}
Z. Qiao, S. Yang, W. Feng, W.-K. Tse, J. Ding, Y. Yao, J. Wang, and Q. Niu, \text{Phys. Rev. B} \textbf{82}, 161414(R) (2010).

\bibitem[Wang(2013)]{proposal6}
Z. Wang, Z. Liu, and F. Liu, \text{Phys. Rev. Lett.} \textbf{110}, 196801 (2013).

\bibitem[Garrity(2013)]{proposal7}
K. F. Garrity, and D. Vanderbilt, \text{Phys. Rev. Lett.} \textbf{110}, 116802 (2013).

\bibitem[Hu(2015)]{proposal8}
J. Hu, Z. Zhu, and R. Wu,  \text{Nano Lett.} \textbf{15}, 2074 (2015).

\bibitem[Fang(2014)]{proposal9}
C. Fang, M. J. Gilbert, and B. A. Bernevig, \text{Phys. Rev. Lett.} \textbf{112}, 046801 (2014).

\bibitem[Wang(2013)]{proposal10}
J. Wang, B. Lian, H. Zhang, Y. Xu, and S.-C. Zhang, \text{Phys. Rev. Lett.} \textbf{111}, 136801 (2013).

\bibitem[Zhang(2014)]{proposal11}
G. Zhang, Y. Li, and C. Wu, \text{Phys. Rev. B} \textbf{90}, 075114 (2014).

\bibitem[Lu(2013)]{proposal12}
H. Lu, A. Zhao, and S. Shen, \text{Phys. Rev. Lett.} \textbf{111}, 146802 (2013).

\bibitem[Liu(2008)]{proposal13}
C. Liu, X. Qi, X. Dai, Z. Fang, and S.-C. Zhang, \text{Phys. Rev. Lett.} \textbf{101}, 146802 (2008).

\bibitem[Ding(2011)]{proposal14}
J. Ding, Z. Qiao, W. Feng, Y. Yao, and Q. Niu, \text{Phys. Rev. B} \textbf{84}, 195444 (2011).

\bibitem[Ezawa(2012)]{proposal15}
M. Ezawa, \text{Phys. Rev. Lett.} \textbf{109}, 055502 (2012).

\bibitem[Zhang(2012)]{proposal16}
H. Zhang, C. Lazo, S. Blügel, S. Heinze, and Y. Mokrousov, \text{Phys. Rev. Lett.} \textbf{108}, 056802 (2012).

\bibitem[(Qiao2014)]{proposal17}
Z. Qiao, W. Ren, H. Chen, L. Bellaiche, Z. Zhang, A. H. MacDonald, and Q. Niu, \text{Phys. Rev. Lett.} \textbf{112}, 116404 (2014).

\bibitem[Wu(2014)]{proposal18}
S.-C. Wu, G. Shan, and B. Yan, \text{Phys. Rev. Lett.} \textbf{113}, 256401 (2014).

\bibitem[Huang(2014)]{proposal19}
S.-M. Huang, S.-T. Lee, and C.-Y. Mou, \text{Phys. Rev. B} \textbf{89}, 195444 (2014).

\bibitem[Pan(2014)]{proposal20}
H. Pan, Z. Li, C.-C. Liu, G. Zhu, Z. Qiao, and Y. Yao, \text{Phys. Rev. Lett.} \textbf{112}, 106802 (2014).

\bibitem[Zhao(2020)]{QAH-Ctunable1}
Y. Zhao, R. Zhang, R. Mei, L. Zhou, H. Yi, Y. Zhang, J. Yu, R. Xiao, K. Wang, N. Samarth, M. H. W. Chan, C. Liu, and C.-Z. Chang, \text{Nature} \textbf{588}, 419 (2020).

\bibitem[Li(2022)]{QAH-Ctunable2}
Z. Li, Y. Han, and Z. Qiao, \text{Phys. Rev. Lett.} \textbf{129}, 036801 (2022).

\bibitem[Yang(2022)]{Yang2022_1}
J.-E. Yang, and H. Xie, \text{Front. Phys.} \textbf{17}, 63504 (2022).

\bibitem[Han(2023)]{QAH-graphene_CrI3}
Y. Han, Z. Yan, Z. Li, X. Xu, Z. Zhang, Q. Niu, and Z. Qiao, \text{Phys. Rev. B} \textbf{107}, 205412 (2023).

\bibitem[Li(2023)]{Li2023}
L. Li, M. Wu, and X. Lu, \text{Front. Phys.} \textbf{18}, 43401 (2023).

\bibitem[Wang(2023)]{Wang_2023}
Y. Wang, F. Zhang, M. Zeng, H. Sun, Z. Hao, Y. Cai, H. Rong, C. Zhang, C. Liu, X. Ma, L. Wang, S. Guo, J. Lin, Q. Liu, C. Liu, and C. Chen,  \text{Front. Phys.} \textbf{18}, 21304 (2023).

\bibitem[Deng(2024)]{QAH-Ctunable3}
P. Deng, Y. Han, P. Zhang, S. K. Chong, Z. Qiao, and K. L. Wang, \text{Phys. Rev. B} \textbf{109}, L201402 (2024).

\bibitem[Li(2024)]{Li2024}
S. Li, K. Wei, Q. Liu, Y. Tang, and T. Jiang, \text{Front. Phys.} \textbf{19}, 42501 (2024).

\bibitem[Xiao(2007)]{QVH1}
D. Xiao, W. Yao, and Q. Niu, \text{Phys. Rev. Lett.} \textbf{99}, 236809 (2007).

\bibitem[Martin(2008)]{QVH2}
I. Martin, Y. M. Blanter, and A. F. Morpurgo, \text{Phys. Rev. Lett.} \textbf{100}, 036804 (2008).

\bibitem[Gorbachev(2014)]{QVH3}
R. V. Gorbachev, J. C. W. Song, G. L. Yu, A. V. Kretinin, F. Withers, Y. Cao, A. Mishchenko, I. V. Grigorieva, K. S. Novoselov, L. S. Levitov, and A. K. Geim, \text{Science}, \textbf{346}, 448 (2014).

\bibitem[Qiao(2011)]{QVH_C}
Z. Qiao, W.-K. Tse, H. Jiang, Y. Yao, and Q. Niu, \text{Phys. Rev. Lett.} \textbf{107}, 256801 (2011).

\bibitem[Zeng(2022)]{Zeng_2022}
J. Zeng, R. Xue, T. Hou, Y. Han, and Z. Qiao, \text{Front. Phys.}  \textbf{17}, 63503 (2022).

\bibitem[Zheng(2023)]{Zheng_2023}
G. Zheng, S. Qu, W. Zhou, and F. Ouyang, \text{Front. Phys.} \textbf{18}, 53302 (2023).

\bibitem[Guo(2024)]{Guo_2024}
S.-D. Guo, Y.-L. Tao, G. Wang, S. Chen, D. Huang, and Y. S. Ang, \text{Front. Phys.} \textbf{19}, 23302 (2024).

\bibitem[Tang(2024)]{Tang_2024}
J. Tang, S. Wang, and H. Yu, \text{Front. Phys.} \textbf{19}, 43210 (2024).

\bibitem{Xun_2024}
W. Xun, C. Wu, H. Sun, W. Zhang, Y. Wu, P. Li, Nano Lett. \textbf{24}, 3541 (2024).

\bibitem{Li_2023_2}
P. Li, C. Wu, C. Peng, M. Yang, and W. Xun, Phys. Rev. B \textbf{108}, 195424 (2023).

\bibitem{Li_2024}
P. Li, B. Liu, S. Chen, W. Zhang, and Z. Guo, Chinese Phys. B \textbf{33}, 017505 (2024).


\bibitem[Pan(2015)]{VP-QAH-disorder}
H. Pan, X. Li, H. Jiang, Y. Yao, and S. A. Yang, \text{Phys. Rev. B} \textbf{91}, 045404 (2015).

\bibitem[Yu(2015)]{VP-QAH-electronic}
Z. Yu, H. Pan, and Y. Yao, \text{Phys. Rev. B} \textbf{92}, 155419 (2015).

\bibitem[Liu(2015)]{half-H-Bi bilayer}
C.-C. Liu, J.-J. Zhou, and Y. Yao, \text{Phys. Rev. B} \textbf{91}, 165430 (2015).

\bibitem[Niu(2015)]{VP-H-Bilayer}
C. Niu, G. Bihlmayer, H. Zhang, D. Wortmann, S. Blügel, and Y. Mokrousov, \text{Phys. Rev. B} \textbf{91}, 041303(R) (2015).

\bibitem[Zhou(2017)]{VP-ferri}
J. Zhou, Q. Sun, and P. Jena, \text{Phys. Rev. Lett.} \textbf{119}, 046403 (2017).

\bibitem[Zhang(2020)]{VP-MX2}
H. Zhang, W. Yang, Y. Ning, and X. Xu, \text{Phys. Rev. B} \textbf{101}, 205404 (2020).

\bibitem[Ai(2021)]{VP-MoSiX}
H. Ai, D. Liu, J. Geng, S. Wang, K. H. Lo, and H. Pan, \text{Phys. Chem. Chem. Phys.} \textbf{23}, 3144 (2021).

\bibitem[Rehman(2022)]{Valley-symmetry-broken}
M. U. Rehman, Z. Qiao, and J. Wang, \text{Phys. Rev. B} \textbf{105}, 165417 (2022).

\bibitem[Zhan(2022)]{VP+QAH1}
F. Zhan, Z. Ning, L.-Y. Gan, B. Zheng, J. Fan, and R. Wang, \text{Phys. Rev. B} \textbf{105}, L081115 (2022).

\bibitem[Xie(2022)]{VP+QAH2}
Y.-M. Xie, C.-P. Zhang, J.-X. Hu, K. F. Mak, and K. T. Law, \text{Phys. Rev. Lett.} \textbf{128}, 026402 (2022).

\bibitem[Zhan(2023)]{VP-QAH-Ctunable3}
F. Zhan, J. Zeng, Z. Chen, X. Jin, J. Fan, T. Chen, and R. Wang, \text{Nano Lett.} \textbf{23}, 2166 (2023).

\bibitem[Guo(2023)]{Guo_2023}
S.-D. Guo, Y.-L. Tao, W.-Q. Mu, and B.-G. Liu, \text{Front. Phys.} \textbf{18}, 33304 (2023).

\bibitem{Li_2023}
P. Li, X. Yang, Q. Jiang, Y. Wu, and W. Xun, Phys. Rev. Materials \textbf{7}, 064002 (2023).

\bibitem[Marrazzo(2018)]{PHS1}
A. Marrazzo, M. Gibertini, D. Campi, N. Mounet, and N. Marzari, \text{Phys. Rev. Lett.} \textbf{120}, 117701 (2018).

\bibitem[ Marrazzo(2019)]{PHS2}
A. Marrazzo, M. Gibertini, D. Campi, N. Mounet, and N. Marzari, \text{Nano Lett.} \textbf{19}, 8431 (2019).

\bibitem[Cucchi(2020)]{PHS experiment}
I. Cucchi, A. Marrazzo, E. Cappelli, S. Riccò, F. Y. Bruno, S. Lisi, M. Hoesch, T. K. Kim, C. Cacho, C. Besnard, E. Giannini, N. Marzari, M. Gibertini, F. Baumberger, and A. Tamai, \text{Phys. Rev. Lett.} \textbf{124}, 106402 (2020).

\bibitem{Liu2021}
Z. Liu, Y. Han, Y. Ren, Q. Niu, and Z. Qiao, Phys. Rev. B \textbf{104}, L121403 (2021).

\bibitem{Zhu2023}
X. Zhu, Y. Chen, Z. Liu, Y. Han, and Z. Qiao, Front. Phys. \textbf{18} 23302 (2023).

\bibitem{Kane_2005}
C. L. Kane and E. J. Mele, Phys. Rev. Lett. \textbf{95}, 226801 (2005).

\bibitem{Xiao2010}
D. Xiao, M.-C. Chang, and Q. Niu, Rev. Mod. Phys. \textbf{82}, 1959 (2010).

\bibitem{Ceresoli_2006}
D. Ceresoli, T. Thonhauser, D. Vanderbilt, and R. Resta, Phys. Rev. B \textbf{74}, 024408 (2006).

\bibitem{SM}
See Supplemental Material at [XXXURL] for more information about the origin of difference in phase diagrams between points $B$ and $C$, the two contributions of orbital magnetization and the analysis of slope $dM_{z}/d\mu$.

\bibitem{Liu_2011}
C.-C. Liu, H. Jiang, and Y. Yao, Phys. Rev. B \textbf{84}, 195430 (2011).

\bibitem{Wang_2015}
P. Wang, L. Lu, and K. Bertoldi, Phys. Rev. Lett. \textbf{115}, 104302 (2015).

\bibitem{Mousavi_2015}
S. Mousavi, A. Khanikaev, and Z. Wang, Nat. Commun. \textbf{6}, 8682 (2015).

\bibitem{Khanikaev_2017}
A. B. Khanikaev and G. Shvets, Nat. Photonics \textbf{11}, 763 (2017).

\bibitem{Mittal_2019}
S. Mittal, V. V. Orre. D. Leykam, Y. Chong, and M. Hafezi, Phys. Rev. Lett. \textbf{123}, 043201 (2019).



\end{thebibliography}
\end{document}